# MQTE: A Measurement-Based Quantum Algorithm for Robust Energy Spectrum Estimation in the NISQ Era

Qing-Xing Xie[a], Yong-Kang Duan[a], Fa-Hui Liu[a], Yan Zhao[bc]*

[a] *Department of Physics, Hubei University, Wuhan 430062, P. R. China.*

[b] *College of Materials Science and Engineering, Sichuan University, Chengdu 610065, P. R. China.*

[c] *The Institute of Technological Sciences, Wuhan University, Wuhan 430072, P. R. China.*

*yanzhao@scu.edu.cn

**Abstract**

Extracting energy spectra from quantum Hamiltonians is a fundamental task for quantum simulation, yet remains challenging on noisy intermediate-scale quantum (NISQ) devices. We propose Measured Quantum Time Evolution (MQTE), an ancilla-free algorithm that estimates energy gaps by applying real-time evolution to a reference state and measuring time-resolved probabilities via repeated projective measurements. Spectral analysis of these signals reveals oscillation frequencies corresponding to eigenvalue differences. Crucially, MQTE exhibits inherent robustness to quantum hardware noise and sampling errors: these disturbances manifest as a white-noise background, which does not distort the underlying spectral structure but rather obscures the frequency information. By increasing the number of measurement samples, the intensity of the background white noise can be suppressed, thereby recovering the original spectral content. We validate the algorithm's performance via numerical simulations on one- and two-dimensional Heisenberg models, demonstrating accurate extraction of energy gaps and resilience against both sampling and circuit-level noise. Experimental implementation on the superconducting quantum processor Tianyan-176-II further confirms the practical feasibility and noise tolerance of MQTE under real hardware conditions. This work provides a robust and scalable framework for quantum spectral estimation in the NISQ era.

## 1. Introduction

Quantum simulation of physical systems is one of the most promising applications of quantum computing, with the potential to unlock insights into complex quantum phenomena that are beyond the reach of classical methods[1–4]. A central challenge in such simulations is the efficient and accurate extraction of spectral information—particularly eigenenergies and energy gaps—of a given Hamiltonian, which determine ground-state properties, excitation spectra, and dynamical behavior[5–9]. While algorithms such as the Variational Quantum Eigensolver (VQE)[10–13] and Quantum Phase Estimation (QPE)[14–16] have laid foundational groundwork, their practical deployment on current hardware remains limited: VQE is prone to barren plateaus and optimization convergence issues, while QPE requires deep circuits and high-fidelity gates beyond the reach of present-day devices.

In this work, we introduce Measured Quantum Time Evolution (MQTE), a quantum algorithm for probing the energy spectrum of a Hamiltonian through real-time dynamics. MQTE operates entirely without ancillary qubits: it applies the real-time evolution operator of the Hamiltonian directly onto a reference state and performs repeated projective measurements at different evolution times. The resulting time-resolved measurement probabilities exhibit oscillatory dynamics, whose frequencies are determined by the energy differences between eigenstates that have non-zero overlap with the initial reference state. By applying spectral analysis to these probability signals, MQTE enables the reconstruction of energy gaps—even when individual eigenenergies remain inaccessible—making it particularly well-suited for studying excitation spectra[17,18] and many-body dynamics[19–21].

Crucially, MQTE exhibits strong resilience to quantum hardware noise—a critical advantage in the Noisy Intermediate-Scale Quantum(NISQ)[22–25] era. Remarkably, its robustness improves with increasing measurement sampling: higher shot counts enhance signal clarity and suppress the impact of gate noise and measurement errors, leading to more stable spectral estimates. This "sampling-enhanced robustness" arises from the fundamental nature of dominant noise sources—both circuit-level decoherence and measurement imperfections typically manifest as random perturbations that affect individual qubit outcomes with approximately equal probability for collapsing into $|0\rangle$ or $|1\rangle$. In contrast, the noise-free time-evolved state exhibits non-uniform, structured collapse probabilities whose temporal dynamics are governed by coherent oscillations imprinted by the Hamiltonian's eigenenergies.

As a result, hardware noise primarily adds a white-noise background to the measured time-domain signal, while preserving the underlying oscillatory frequencies that encode energy differences. These characteristic frequencies—though initially masked by statistical fluctuations—can be reliably recovered through spectral analysis as the number of measurements increases. Ensemble averaging progressively suppresses noise-induced fluctuations, allowing the structured energy spectrum to emerge clearly. This separation between coherent quantum dynamics and unstructured noise endows MQTE with an inherent fault-tolerance-like behavior, making it especially suitable for near-term quantum devices where high-fidelity control remains challenging.

In this paper, we propose a spectral perspective on quantum simulation: rather than extracting physical information directly from the final quantum state, MQTE infers eigenenergy information from the frequency content of time-resolved measurement signals. This approach exploits a key structural distinction—coherent dynamics generate sharp spectral peaks corresponding to energy differences, while quantum hardware noise (e.g., decoherence and readout errors) contributes broadband, approximately white fluctuations. Consequently, increasing the number of measurement samples enables effective noise suppression through classical post-processing, without requiring quantum error correction.

The remainder of this paper is organized as follows. In Section 2, we establish the theoretical foundation of the MQTE protocol and detail its circuit implementation. In Section 3, we demonstrate its performance through numerical simulations on simple one- and two-dimensional Heisenberg models, showcasing its accuracy in extracting energy gaps as well as its robustness against both sampling errors and circuit-level noise. In Section 4, we present experimental results obtained by implementing MQTE on real NISQ-era quantum hardware, providing empirical validation of its practical feasibility and noise resilience. Finally, Section 5 concludes the paper and outlines promising directions for future work.

## 2. Theory and Methodology

### 2.1 Algorithms Theory

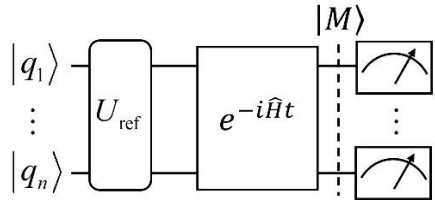

**Figure 1.** Quantum circuits for the Measured Quantum Time Evolution (MQTE) algorithms. The system register, consisting of $q_1 \cdots q_n$ qubits, is initialized in a reference state $|\psi_{\text{ref}}\rangle$. The time-evolution operator $e^{-i\hat{H}t}$ is applied directly to this register, followed by projective measurements in the computational basis.

The MQTE protocol operates without any ancillary qubits. As illustrated in Figure 1, the system register is first prepared in a reference state $|\psi_{\text{ref}}\rangle$ via a unitary operator $U_{\text{ref}}$. This reference state is chosen based on physical intuition or prior knowledge. After state preparation, the real-time evolution operator $e^{-i\hat{H}t}$ is applied directly to the system, yielding the time-evolved state:

$$|M(t)\rangle = e^{-i\hat{H}t}|\psi_{\text{ref}}\rangle \tag{1}$$

Projective measurements are then performed in the computational basis at multiple time points $t$. Denoting the measurement outcome corresponding to configuration $|\varphi_i\rangle$, the probability of observing this outcome at time $t$ is:

$$P_{ij}(t) = \left|\langle\varphi_i|e^{-i\hat{H}t}|\varphi_j\rangle\right|^2$$

where we have taken $|\psi_{\text{ref}}\rangle = |\varphi_j\rangle$ for concreteness.

Assuming the Hamiltonian $\hat{H}$ has eigenstates $|\Psi_k\rangle$ with eigenvalues $E_k$, we expand the reference and measurement configurations in this eigenbasis:

$$|\varphi_j\rangle = \sum_k c_{kj}|\Psi_k\rangle \quad c_{kj} = \langle\Psi_k|\varphi_j\rangle \tag{2}$$

Substituting into the expression for $P_{ij}(t)$, we obtain:

$$\begin{aligned}
P_{ij}(t) &= \left|\sum_k c_{kj}\langle\varphi_i|e^{-i\hat{H}t}|\Psi_k\rangle\right|^2 \\
&= \left|\sum_k c_{kj}e^{-iE_kt}c_{ki}^*\right|^2 = \sum_{k'} c_{k'j}^* e^{iE_{k'}t}c_{k'i} \cdot \sum_k c_{kj}e^{-iE_kt}c_{ki}^* \\
&= \sum_{k'=k} c_{k'j}^*c_{k'i}c_{kj}c_{ki}^* e^{-i(E_k-E_{k'})t} + \sum_{k'>k}\cdots + \sum_{k'<k}\cdots \\
&= \sum_k |c_{ki}|^2|c_{kj}|^2 + \sum_{k>k'}\left[c_{k'j}^*c_{k'i}c_{kj}c_{ki}^* \cdot e^{-i(E_k-E_{k'})t} + \text{c.c.}\right] \\
&= \sum_k |c_{ki}|^2|c_{kj}|^2 + \sum_{k>k'} 2\text{Re}\left[c_{k'j}^*c_{k'i}c_{kj}c_{ki}^* \cdot e^{-i(E_k-E_{k'})t}\right]
\end{aligned} \tag{3}$$

where the three summation terms in the third line share an identical functional form, differing only in their summation ranges; for brevity, the common summand $c_{k'j}^*c_{k'i}c_{kj}c_{ki}^*e^{-i(E_k-E_{k'})t}$ is denoted by "⋯" in the latter two terms. The overlap coefficients $c_{kj} = \langle\Psi_k|\varphi_j\rangle$ between eigenstates $|\Psi_k\rangle$ and a configurations $|\varphi_j\rangle$ are generally complex. However, for time-reversal-invariant systems described with real orbitals—as in standard non-relativistic quantum chemistry calculations—both $|\Psi_k\rangle$ and $|\varphi_j\rangle$ can be chosen as real functions, making all $c_{kj}$ real. Therefore, in the final equality of the above derivation, we have assumed that the coefficients $c_{kj}$ (and similarly $c_{ki}$) are real. Under this physically relevant assumption, Eq. (3) simplifies to:

$$P_{ij}(t) = \sum_{k} c_{ki}^2 c_{kj}^2 + \sum_{k>k'} 2c_{k'j}c_{k'i}c_{kj}c_{ki} \cdot \cos[(E_k - E_{k'})t] \quad (4)$$

Eq. (2) reveals that the time-dependent measurement probabilities consist of a constant background plus oscillatory terms whose frequencies are precisely the energy differences $\Delta E = E_k - E_{k'}$ between pairs of eigenstates that overlap with both $|\varphi_i\rangle$ and $|\varphi_j\rangle$. Thus, by performing spectral analysis—such as a Fourier transform—on the sampled signal $P_{ij}(t)$, one can extract the set of resolvable energy gaps from the positions of spectral peaks, while their amplitudes encode information about the overlaps among the reference state, measurement basis, and true eigenstates.

In practice, time is discretized: let the sampling interval be $\Delta$, and collect N samples at times $t_n = n\Delta$ for $n = 0,1,2\cdots N$. Due to the even symmetry of $P_{ij}(t)$ (i.e., $P_{ij}(t) = P_{ij}(-t)$), the discrete signal can be extended symmetrically to $n = -N, -N+1, \cdots, N-1, N$. The discrete Fourier transform (DFT) of the measured sequence $p_{ij}(n) = P_{ij}(n\Delta)$ is defined as:

$$F(k) = \frac{1}{2N+1}\sum_{n=-N}^{N} p_{ij}(n) e^{-i\frac{2\pi}{2N+1}kn} \quad (5)$$

with inverse transform:

$$p_{ij}(n) = \sum_{k=-N}^{N} F(k) e^{i\frac{2\pi}{2N+1}nk} \quad (6)$$

Because $p_{ij}(n)$ is real and even, $F(k)$ is also real and even. Consequently, the signal can be expressed as:

$$\begin{aligned}
p_{ij}(n) &= F(0) + \sum_{k=1}^{N} F(k) e^{i\frac{2\pi}{2N+1}nk} + \sum_{k=-1}^{-N} F(k) e^{i\frac{2\pi}{2N+1}nk} \\
&= F(0) + \sum_{k=1}^{N} F(k) e^{i\frac{2\pi}{2N+1}nk} + \sum_{k=1}^{N} F(-k) e^{-i\frac{2\pi}{2N+1}nk} \\
&= F(0) + \sum_{k=1}^{N} F(k) \left( e^{i\frac{2\pi}{2N+1}nk} + e^{-i\frac{2\pi}{2N+1}nk} \right) \\
&= F(0) + \sum_{k=1}^{N} 2F(k) \cdot \cos\left(\frac{2\pi}{(2N+1)\Delta} k \cdot n\Delta\right)
\end{aligned} \quad (7)$$

Comparing Eq. (7) with Eq. (4), we identify the correspondence between Fourier frequencies and physical energy differences:

$$\begin{aligned}
|E_m - E_n| &= \frac{2\pi}{(2N+1)\Delta} k \\
c_{nj}c_{ni}c_{mj}c_{mi} &= F(k)
\end{aligned} \quad (8)$$

Thus, the full set of resolvable energy gaps—and associated transition amplitudes—can be reconstructed from the DFT spectrum. The frequency resolution is limited by the maximum observation time $T = N\Delta$; Given that $N \gg 1$, the smallest resolvable energy difference is $\delta E_{min} = \pi/T$. To achieve a target spectral accuracy $\varepsilon > 0$, the number of time samples is chosen as $N = \left\lceil \frac{\pi}{\Delta\varepsilon} \right\rceil$ to ensure that the frequency resolution $\frac{2\pi}{(2N+1)\Delta} \leq \varepsilon$.

A key advantage of MQTE is its hardware efficiency: it requires no ancilla qubits and avoids controlled-unitary operations, significantly reducing circuit depth and gate count compared to interferometric approaches. This makes MQTE particularly well-suited for implementation on current NISQ devices, where coherence times and gate fidelities remain limited.

## 2.2 Algorithms Implementation

To facilitate implementation, we summarize the execution procedure of the Measured Quantum Time Evolution (MQTE) algorithm in the following pseudocode.

---

**Algorithm**: Measured Quantum Time Evolution (MQTE)

**Input:**
    Hamiltonian $\hat{H}$
    Computational accuracy $\varepsilon$

**Output:**
    Eigen-energie differences $\{|E_m - E_n|\}$
    Wavefunction overlaps: $\{c_{nj}c_{ni}c_{mj}c_{mi}\}$

Select an appropriate reference state $|\varphi_j\rangle$

Design $U_{\text{ref}}$ quantum circuits based on the selected reference state $|\varphi_j\rangle$

Set an appropriate sampling interval $\Delta$

Calculate the number of time samples $N = \left\lceil \frac{\pi}{\Delta\varepsilon} \right\rceil$

Initialize a collection of arrays $\{p_{ij}\}$, each of length $2N + 1$

**for** $n = 0$ to $N$ **do**
    set $t = n \cdot \Delta$
    Initialize target qubits to $|0\rangle$
    Apply $U_{\text{ref}}$ circuit on target qubits to prepare $|\varphi_j\rangle$

---

>       Apply $e^{-i\hat{H}t}$ circuit acting on target qubits
>
>       Perform repeated projective measurements on the output state to estimate the probability $P_{ij}(t)$ of observing the system in configuration $|\varphi_i\rangle$
>       **for** $i$ in each detected $|\varphi_i\rangle$ state **do**
>             Store $P_{ij}(t)$ in $p_{ij}[n]$
>       **end for**
> **end for**
> **for** $n = N + 1$ to $2N$ **do**
>       **for** $i$ in each detected $|\varphi_i\rangle$ state **do**
>             set $p_{ij}[n] = p_{ij}[2N + 1 - n]$
>       **end for**
> **end for**
> **for** $i$ in each detected $|\varphi_i\rangle$ state **do**
>       Perform DFT in Eq. (5) on the sequence $p_{ij}[n]$ to obtain its frequency spectrum $G(k)$
>       Analyze spectrum, find each peak:
>
>       $|E_m - E_n| = \frac{2\pi}{(2N+1)\Delta} k$
>
>       $c_{nj}c_{ni}c_{mj}c_{mi} = F(k)$
>
> Return $\{|E_m - E_n|\}$ $\{c_{nj}c_{ni}c_{mj}c_{mi}\}$

## 3. Numerical simulation and discussion

### 3.1 Algorithm effectiveness verification

To evaluate the performance of the Measured Quantum Time Evolution (MQTE) algorithm, we carry out numerical simulations using the MindSpore Quantum framework[26], which provides a powerful environment for quantum computation simulation. We consider two canonical many-body systems: a one-dimensional (1D) Heisenberg chain with 10 sites and a two-dimensional (2D) Heisenberg model on a 3×3 square lattice. The Hamiltonian for both models is given by:

$$\hat{H} = J \sum_{\langle j,k \rangle} [\sigma_j^x \cdot \sigma_k^x + \sigma_j^y \cdot \sigma_k^y] + h \sum_{\langle j,k \rangle} \sigma_j^z \cdot \sigma_k^z \qquad (9)$$

where $J$ denotes the spin-exchange coupling in the $x$-$y$ plane, $h$ is the Ising-type coupling along the $z$-axis, and $\langle j, k \rangle$ indicates summation over nearest-neighbor pairs. The operators $\sigma_j^\alpha (\alpha = x, y, x)$ are the Pauli matrices acting on site $j$. Both models employ open boundary conditions, and we set $J = 1$ and $h = 2$ throughout.

The reference state is chosen as the Néel state—i.e., $|\uparrow\downarrow\uparrow\downarrow \cdots \rangle$ in 1D and a checkerboard antiferromagnetic configuration in 2D—which exhibits significant overlap with low-lying excited states in antiferromagnetic systems. Time evolution under $e^{-i\hat{H}t}$ is implemented via second-order Trotter–Suzuki decomposition[27–30] with a fixed Trotter step $\tau = 0.01$, ensuring that discretization errors remain well-controlled across all simulated evolution times.

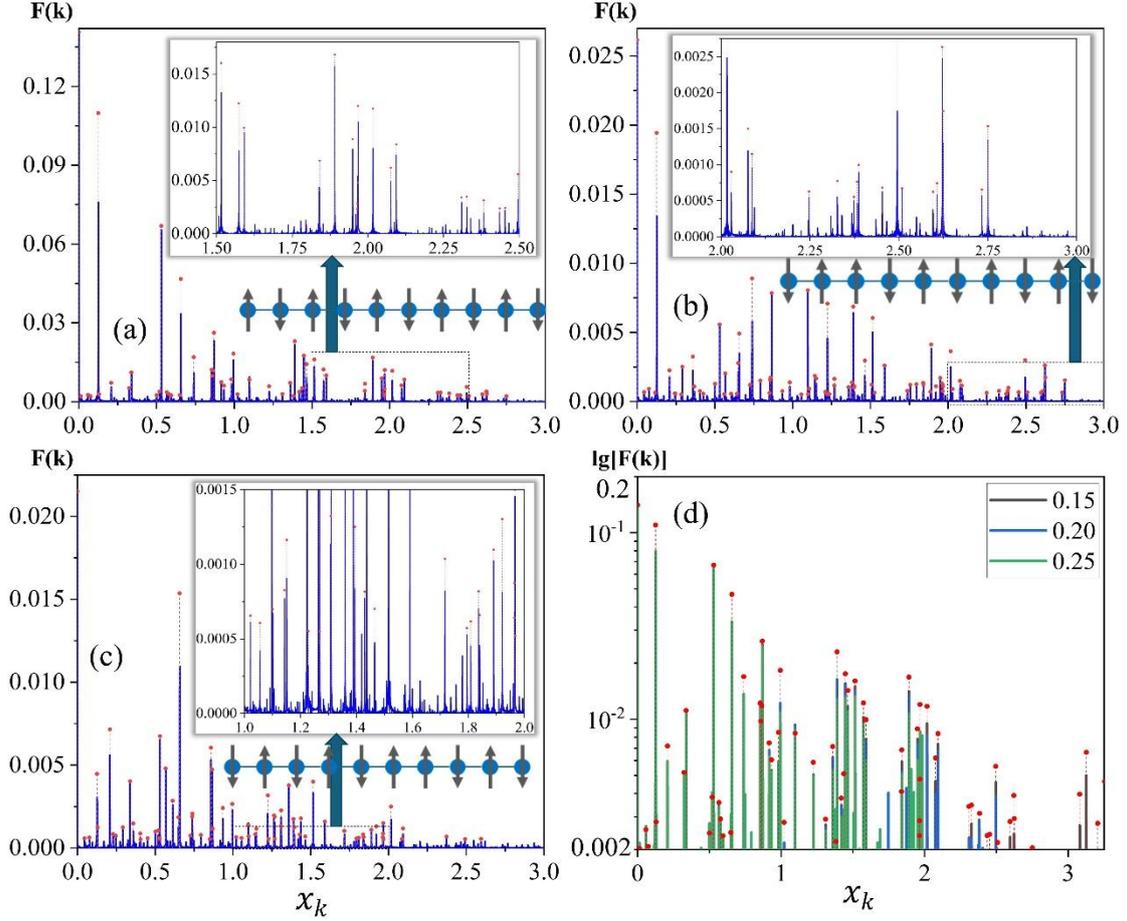

Figure 2. DFT spectra of the time-resolved measurement probabilities $p_{ij}(n)$ for a 10-site 1D Heisenberg model with open boundaries, computed using the MQTE algorithm. The reference state $|\varphi_j\rangle$ is initialized as the Néel state shown in panel (a). Panels (a), (b), and (c) display the DFT spectra corresponding to projective measurements onto different configurations $|\varphi_i\rangle$, represented by the spin patterns beneath each plot. The horizontal axis represents the frequency variable $x_k = \frac{k}{(2N+1)\Delta}$, which is proportional to the physical energy differences via $|E_m - E_n| = 2\pi x_k$. The blue lines indicate the simulated spectral amplitudes $F(k)$, while the red dots denote the exact energy gaps obtained from full diagonalization, with heights proportional to $c_{nj}c_{ni}c_{mj}c_{mi}$. For visual clarity, only gap-transitions with spectral weight exceeding a panel-specific threshold are shown: 0.002 for (a), and 0.0005 for both (b) and (c). All simulations are performed with a fixed maximum evolution time $T = N\Delta = 2000$. For panels (a)–(c), the sampling interval is set to $\Delta = 0.1$. Panel (d) compares the DFT spectra under varying sampling

intervals: $\Delta = 0.15$ (black), 0.20 (blue), and 0.25 (green), with all other parameters—including the maximum time $T = 2000$—identical to those in panel (a). The vertical axis in (d) is logarithmic to highlight spectral structure across different resolutions. The inset zooms into regions of high spectral density to emphasize peak alignment with theoretical predictions.

For 10-site 1D Heisenberg model, as shown in Figure 2(a)–(c), the DFT of the time-resolved measurement probabilities $p_{ij}(n) = |\langle\varphi_i|e^{-i\hat{H}\cdot n\Delta}|\varphi_j\rangle|^2$ yields sharp spectral peaks that align precisely with the exact energy gaps (red dot-vertical lines) obtained from full diagonalization. These reference energy differences correspond to $|E_m - E_n|/2\pi$ on the horizontal axis, and their spectral weights are proportional to the product $c_{nj}c_{ni}c_{mj}c_{mi}$, where $c_{\alpha\beta} = \langle\Psi_\alpha|\varphi_\beta\rangle$ denotes the overlap between computational basis states $|\varphi_\beta\rangle$ and exact eigenstates $|\Psi_\alpha\rangle$ of the Hamiltonian. In principle, the full spectrum contains contributions from all pairs of eigenstates; however, only gap-transitions with sufficiently large weight—specific amplitude threshold—are displayed for clarity. This selective visualization highlights the dominant spectral features while suppressing negligible contributions.

The reference state $|\varphi_j\rangle$ is chosen as the Néel configuration depicted in panel (a), and measurements are performed in different computational basis states $|\varphi_i\rangle$, each corresponding to a distinct spin pattern. Despite the simplicity of the measurement protocol, MQTE successfully resolves multiple low-lying excitation energies, confirming its ability to extract meaningful spectral information from local observables. Comparing the spectra in panels (a), (b), and (c), it is evident that despite variations in the spectral structure, the positions of the spectral peaks remain largely consistent across different panels. This consistency arises because all simulations model the same physical system, hence they share identical eigenenergy gaps. The only difference lies in the projective measurements onto distinct computational basis states $|\varphi_i\rangle$, which result in different collapsed states after measurement. Different $|\varphi_i\rangle$ states have varying overlaps with the eigenstates, leading to different spectral amplitudes for the same energy gap.

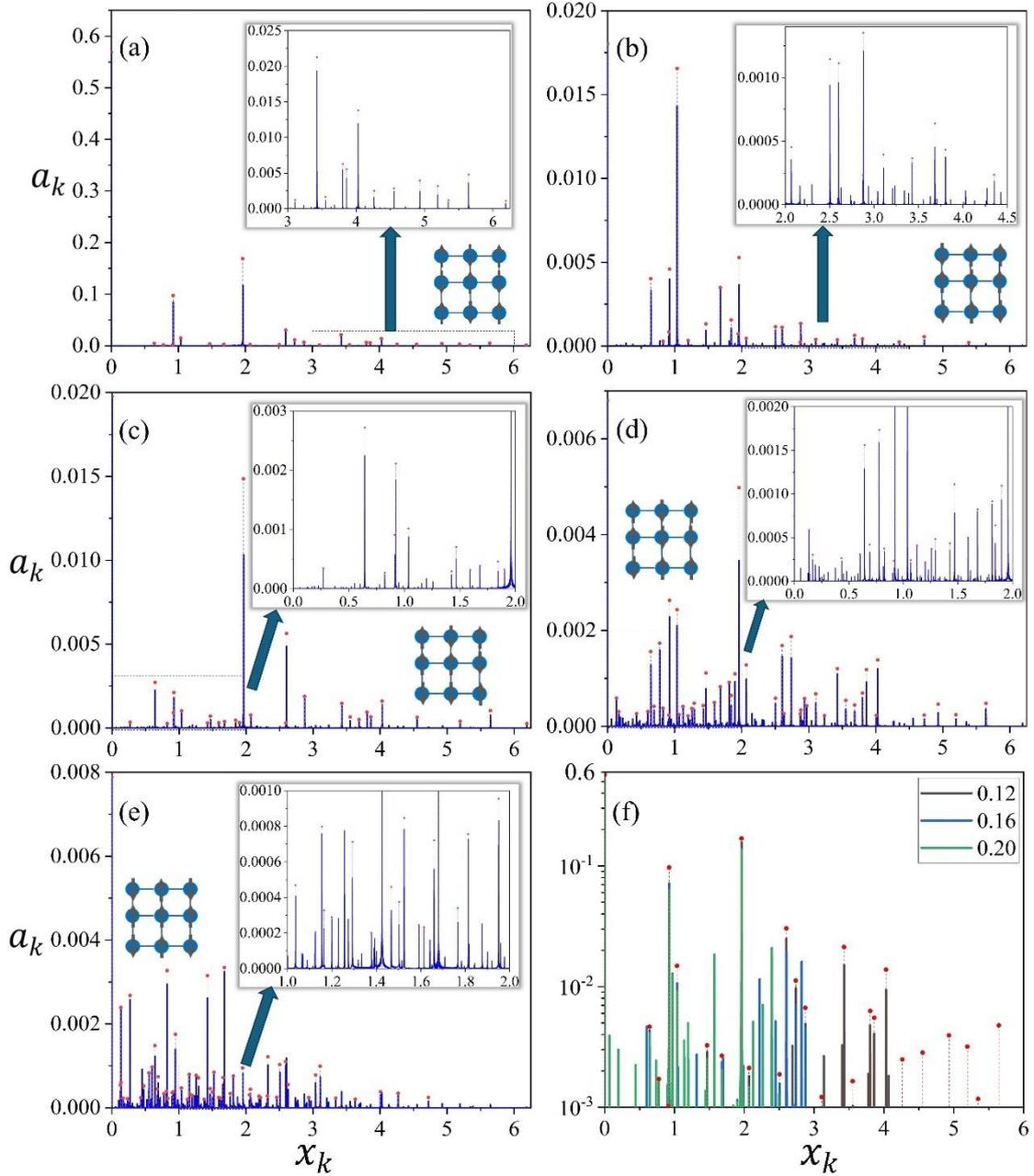

Figure 3. Similar to Figure 2, panels (a)–(e) illustrate the DFT spectra of $p_{ij}(n)$ for a 3×3 two-dimensional Heisenberg model with open boundary conditions, computed using the MQTE algorithm. The reference state $|\varphi_j\rangle$ is the Néel configuration depicted in panel (a), and all simulations employ a fixed sampling interval $\Delta = 0.08$. The blue lines represent simulated spectral amplitudes, while red vertical lines indicate exact energy gaps obtained from full diagonalization. Panel (f) repeats the same calculation as in (a)—same reference and measurement states—but with different sampling intervals: $\Delta = 0.12$ (black), $0.16$ (blue), and $0.20$ (green). All simulations use a maximum evolution time $T = N\Delta = 2000$.

Critically, the fidelity of spectral reconstruction depends on the sampling interval $\Delta$. In Figure 2(d), we compare DFT spectra computed with $\Delta$=0.15, 0.20, and 0.25, while keeping the maximum evolution time fixed at $T = N\Delta = 2000$. As $\Delta$ increases, high-frequency components begin to alias into lower frequencies, leading to spurious peaks. This behavior underscores the necessity of satisfying the Nyquist criterion ($\Delta < \pi/|E_m - E_n|$) to avoid distortion of the reconstructed spectrum. Notably, for $\Delta = 0.1$ (used in panels (a)–(c)), no significant aliasing is observed, and all dominant peaks match theoretical predictions within numerical precision—demonstrating that MQTE achieves accurate spectral estimation when properly sampled in time.

To further validate the generality and scalability of the MQTE algorithm, we extend our numerical study to a 3×3 two-dimensional Heisenberg model with open boundaries, as shown in Figure 3. Similar to the one-dimensional case, the DFT spectra of $p_{ij}(n)$ exhibit sharp peaks that align closely with the exact energy differences (red lines), confirming the algorithm's ability to extract accurate spectral information in higher dimensions. Panels (a)–(e) correspond to different measurement outcomes $|\varphi_i\rangle$, each associated with a distinct spin configuration depicted below the spectrum. Despite the complexity of the 2D system, the spectral features remain well-resolved, and the positions of dominant peaks are consistent across panels—reflecting the fact that all simulations describe the same Hamiltonian and thus share identical eigenenergy gaps. Panel (f) confirms the same conclusion as in the 1D case: only when the sampling interval satisfies $\Delta < \pi/|E_m - E_n|$ for all relevant eigenenergy pairs can the $p_{ij}(n)$ spectrum faithfully reproduce the true energy gaps. Otherwise, large-gap transitions alias into lower frequencies, distorting the spectrum.

It is important to note that, in principle, the ideal spectral weight of a peak at frequency $|E_m - E_n|/2\pi$ is proportional to the product $c_{nj}c_{ni}c_{mj}c_{mi}$, where $c_{\alpha\beta} = \langle\Psi_\alpha|\varphi_\beta\rangle$ quantifies overlaps between computational basis states and eigenstates. However, in practice, the reconstructed DFT amplitude $F(k)$ may deviate from this ideal value due to finite-time effects: even when the sampling interval $\Delta$ satisfies the Nyquist criterion, the finite maximum evolution time $T = N\Delta$ acts as a rectangular window in the time domain, causing spectral leakage that redistributes intensity across neighboring frequencies. While this introduces quantitative errors in estimating the exact overlap magnitudes, the relative heights of dominant peaks remain robust indicators of which eigenstate transitions contribute most significantly to the dynamics. Thus, qualitative conclusions about the composition of the reference state—for instance, identifying dominant low-energy excitations—remain reliable.

## 3.2 Robustness of MQTE to Sampling Error

**Simulation Results**

All the preceding analysis and discussion are predicated on the assumption that the time-dependent probabilities $p_{ij}(n)$ can be sampled with high accuracy. In practice, implementing the MQTE algorithm requires estimating these probabilities by performing projective measurements of the evolved state $e^{-i\hat{H}t}|\varphi_j\rangle$ in various computational basis states $|\varphi_i\rangle$. However, due to the finite number of quantum measurements, the empirical estimates of $p_{ij}(n)$ inevitably suffer from statistical (sampling) errors.

The results reveal several important features. First, the positions of all dominant spectral peaks remain unchanged across different values of $M$, confirming that sampling noise does not introduce systematic shifts in the inferred energy gaps. Second, the relative heights of strong peaks are also stable, indicating that the qualitative structure of the spectrum—i.e., which transitions dominate—is robust against statistical fluctuations. Third, and most notably, a stochastic background resembling white noise is superimposed on the spectrum, with its amplitude decreasing as $M$ increases. As shown in panel (a), when only $M = 10$ measurements are performed per time step, the root-mean-square (RMS) noise floor reaches approximately $2 \times 10^{-3}$. Under these conditions, any spectral feature with amplitude below this level becomes indistinguishable from statistical fluctuations. As $M$ is increased to 40, 160, 640, 2560, and finally 10240, the noise floor drops progressively: roughly halving each time $M$ is quadrupled. This behavior is fully consistent with the central limit theorem, which predicts a standard deviation scaling as $\sigma \propto M^{-1/2}$. The observed trend holds uniformly across panels (a)-(c), despite differences in the underlying signal amplitudes, underscoring the universal nature of sampling-induced uncertainty.

Critically, this analysis demonstrates that sampling noise affects detectability but not accuracy: while weak transitions may be lost in the noise for small $M$, no spurious energy gaps are created, and the locations of genuine peaks remain faithful to the exact spectrum. Consequently, increasing the number of measurements improves the signal-to-noise ratio, enabling the resolution of increasingly subtle excitations without altering the intrinsic spectral content of the system. This property makes MQTE particularly suitable for near-term quantum devices, where measurement budgets are limited but reliable identification of dominant low-energy features is often sufficient for physical insight.

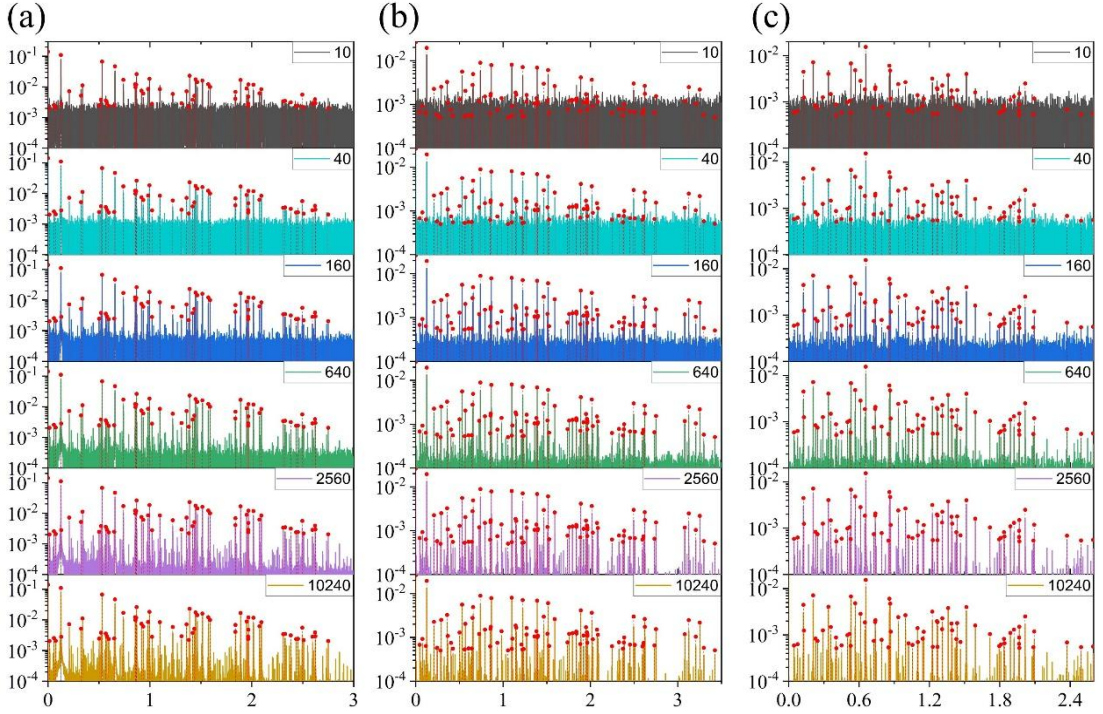

Figure 4. DFT spectra of $p_{ij}(n)$ for the 10-site one-dimensional Heisenberg model, computed with varying numbers of measurement samples (from top to bottom: 10, 40, 160, 640, 2560, and 10240), under the same parameters as in Figure 2(a)–(c). Each panel corresponds to a different measurement basis state $|\varphi_i\rangle$, as labeled in Figure 2. The red dots denote exact energy gaps from full diagonalization.

**Theoretical Analysis**

To gain a deeper understanding of how finite sampling affects the performance of the MQTE algorithm, we develop a statistical model that quantifies the noise introduced by limited measurement repetitions. This framework allows us to predict the noise level in the reconstructed spectrum and establish practical criteria for distinguishing genuine spectral peaks—arising from true energy gaps—from spurious features caused by statistical fluctuations.

Let $p_{ij}(n)$ denote the exact (theoretical) probability of observing the system in computational basis state $|\varphi_i\rangle$ after $n\Delta$ time evolution, starting from reference state $|\varphi_j\rangle$. In an actual simulation or experiment, this probability is estimated from $M$ independent projective measurements at each time step $n$, yielding an empirical estimate $\bar{p}_{ij}(n)$. The deviation between the two is defined as the sampling error:

$$\bar{p}_{ij}(n) = p_{ij}(n) + s_{ij}(n) \tag{10}$$

Since each measurement outcome is a Bernoulli random variable (either success collapse to $|\varphi_i\rangle$, or failure), the estimator $\tilde{p}_{ij}(n)$ follows a binomial distribution with mean $\mathbb{E}[\tilde{p}_{ij}(n)] = p_{ij}(n)$ and variance

$$\text{Var}[\tilde{p}_{ij}(n)] = \frac{p_{ij}(n)[1 - p_{ij}(n)]}{M} \tag{11}$$

By the Central Limit Theorem, for moderate to large $M$, the sampling error $s_{ij}(n)$ is well approximated by a Gaussian distribution:

$$s_{ij}(n) \sim N\left(0, \frac{p_{ij}(n)[1 - p_{ij}(n)]}{M}\right) \tag{12}$$

We now examine how this time-domain noise propagates into the frequency domain. Because the discrete Fourier transform is a linear operation, the spectrum of the noisy signal decomposes as:

$$DFT[\tilde{p}_{ij}(n)] = DFT[p_{ij}(n)] + DFT[s_{ij}(n)] \tag{13}$$

This equation explains a key observation in Figure 4: the positions and relative heights of the dominant spectral peaks remain unchanged across different sampling numbers, while a fluctuating background is superimposed. In other words, sampling noise does not distort the underlying energy spectrum—it merely adds a stochastic component that can obscure weak frequency peaks.

In our simulations, we enforce the symmetry $p_{ij}(n) = p_{ij}(-n)$ (which holds exactly for real Hamiltonians and appropriate initial states), and we construct $\tilde{p}_{ij}(n)$ to preserve this symmetry. Consequently, the error sequence also satisfies $s_{ij}(n) = s_{ij}(-n)$. The DFT coefficient of the sampling noise at frequency index $k$ is then given by:

$$S(k) = \frac{1}{2N+1} \sum_{n=-N}^{N} s_{ij}(n) e^{-i\frac{2\pi}{2N+1}kn} \tag{14}$$

Because each $s_{ij}(n)$ is Gaussian and independent across different $n$, any linear combination—including the above sum—is also Gaussian. Its mean is zero:

$$\mathbb{E}[S(k)] = \frac{1}{2N+1} \sum_{n=-N}^{N} \mathbb{E}[s_{ij}(n)] e^{-i\frac{2\pi}{2N+1}kn} = 0 \tag{15}$$

and its variance is:

$$\begin{aligned} Var[S(k)] &= \mathbb{E}[S^2(k)] - \mathbb{E}^2[S(k)] \\ &= \frac{1}{(2N+1)^2} \mathbb{E}\left[\sum_{n=-N}^{N} s_{ij}(n) e^{-i\frac{2\pi}{2N+1}kn} \times \sum_{m=-N}^{N} s_{ij}(m) e^{-i\frac{2\pi}{2N+1}km}\right] \\ &= \frac{1}{(2N+1)^2} \sum_{n=-N}^{N} \sum_{m=-N}^{N} \mathbb{E}[s_{ij}(n) s_{ij}(m)] e^{-i\frac{2\pi}{2N+1}k(n-m)} \end{aligned} \tag{16}$$

The cross-correlation $\mathbb{E}[s_{ij}(n)s_{ij}(m)]$ vanishes for $n \neq m$ due to statistical independence: $\mathbb{E}[s_{ij}(n)s_{ij}(m)] = \mathbb{E}[s_{ij}(n)] \times \mathbb{E}[s_{ij}(m)] = 0$, and for $n = m$ it equals the variance of $s_{ij}(n)$: $\mathbb{E}[s_{ij}^2(n)] = Var[s_{ij}(n)] + \mathbb{E}^2[s_{ij}(n)] = Var[s_{ij}(n)]$. Thus:

$$\mathbb{E}[s_{ij}(n)s_{ij}(m)] = \begin{cases} \dfrac{p_{ij}(n)[1 - p_{ij}(n)]}{M} & m = n \\ 0 & m \neq n \end{cases} \quad (17)$$

Substituting (17) into (16) simplifies the double sum to a single sum:

$$Var[S(k)] = \frac{1}{(2N+1)^2} \sum_{n=-N}^{N} \frac{p_{ij}(n)[1 - p_{ij}(n)]}{M} \quad (18a)$$

Noting that the function $f(p) = p(1-p)$ attains its maximum value $1/4$ at $p = 1/2$, we obtain a universal upper bound:

$$\sigma_0^2 = p_{ij}(n)[1 - p_{ij}(n)] \leq \frac{1}{4} \text{ for all } n$$

Thus,

$$Var[S(k)] \leq \frac{1}{(2N+1)^2} \cdot \frac{2N+1}{4M} = \frac{1}{4(2N+1)M} \approx \frac{1}{8NM} \quad (18b)$$

where the approximation holds for large $N$.

Therefore, the noise in the frequency domain is a zero-mean Gaussian process with standard deviation bounded by $1/\sqrt{8NM}$. Combining this result with Eq. (7), we find that the sampling noise contributes to the magnitude spectrum as $|2S(k)|$. Consequently, the observed spectral amplitude is the sum of the true signal and a noise term that follows a half-normal distribution:

$$|2S(k)| \sim \text{Half Normal}(0, \sigma^2) \quad \sigma^2 \leq \frac{4\sigma_0^2}{(2N+1)M} \leq \frac{1}{2NM} \quad (19)$$

In practice, when analyzing a sampled spectrum, the probability that a given peak originates from random noise decreases rapidly with its amplitude. For a half-normal distribution, the probability that the noise exceeds $3\sigma$ is approximately 0.27% (i.e., about one fake peak per 370 frequency bins), and the probability of exceeding $4\sigma$ drops to 0.0063% (roughly one fake peak per 15,780 bins). This motivates the use of an amplitude threshold for peak identification: only peaks above a chosen multiple of $\sigma$ are accepted as eigenenergy gaps.

While a $3\sigma$ criterion is commonly adopted in statistical hypothesis testing, it must be applied with caution in spectral analysis due to the large number of frequency bins. In our simulations, the DFT contains approximately 20,000 bins; even with a low per-bin false-alarm rate of 0.27% , we still expect dozens of spurious peaks above $3\sigma$ purely by chance. For instance, in Figure 4(a), where the number of measurements per time step is $M = 10$ and the total number of time steps is $N = 20,000$ , the theoretical upper bound on the noise standard deviation is $\sigma_{\max} = \frac{1}{\sqrt{2NM}} \approx 0.00158$, so $3\sigma_{\max} \approx 0.00474$ . Indeed, a small number of isolated peaks near this level appear in Figure 4(a) at frequencies where no exact energy gap (red vertical lines) exists—strong evidence that they arise from sampling noise. These spurious features can obscure or mimic genuine gaps, complicating accurate eigen-spectral interpretation. To eliminate such false positives, a stricter $4\sigma$ threshold ensures that, on average, fewer than one noise-induced peak appears in the entire spectrum—effectively guaranteeing reliable identification at the cost of potentially missing very weak but real transitions whose amplitudes fall below the threshold.

By contrast, in Figures 4(b) and 4(c), the $3\sigma$ criterion remains highly reliable. Moreover, a direct comparison of panels (a)-(c) reveals that, despite identical sampling parameters ( $M = 10$ ), the overall noise floor differs across panels: it is highest in (a), intermediate in (b), and lowest in (c). This variation stems from the fact that the bound $\sigma_0^2 = 1/4$ is rarely saturated in practice. The actual variance $p_{ij}(n)[1 - p_{ij}(n)]/M$ depends on the specific measurement basis $|\varphi_i\rangle$ : different collapse probabilities $p_{ij}(n)$ lead to different local noise levels, which in turn affect the global noise intensity in the DFT spectrum. Thus, the choice of measurement configuration not only influences which spectral features are visible (via overlap with eigenstates) but also modulates the effective signal-to-noise ratio.

### 3.3 Robustness of MQTE to Circuit Noise
**Simulation Results**

In practical quantum devices, the performance of quantum algorithms is fundamentally limited by various sources of noise arising from imperfect gate operations, environmental coupling, and measurement errors. These imperfections cause deviations in the execution of quantum logic gates, leading to decoherence and the transformation of pure states into mixed states. In the current NISQ era, circuit-level noise is an unavoidable factor, making noise robustness a critical metric for evaluating the feasibility of quantum algorithms on real hardware[31,32].

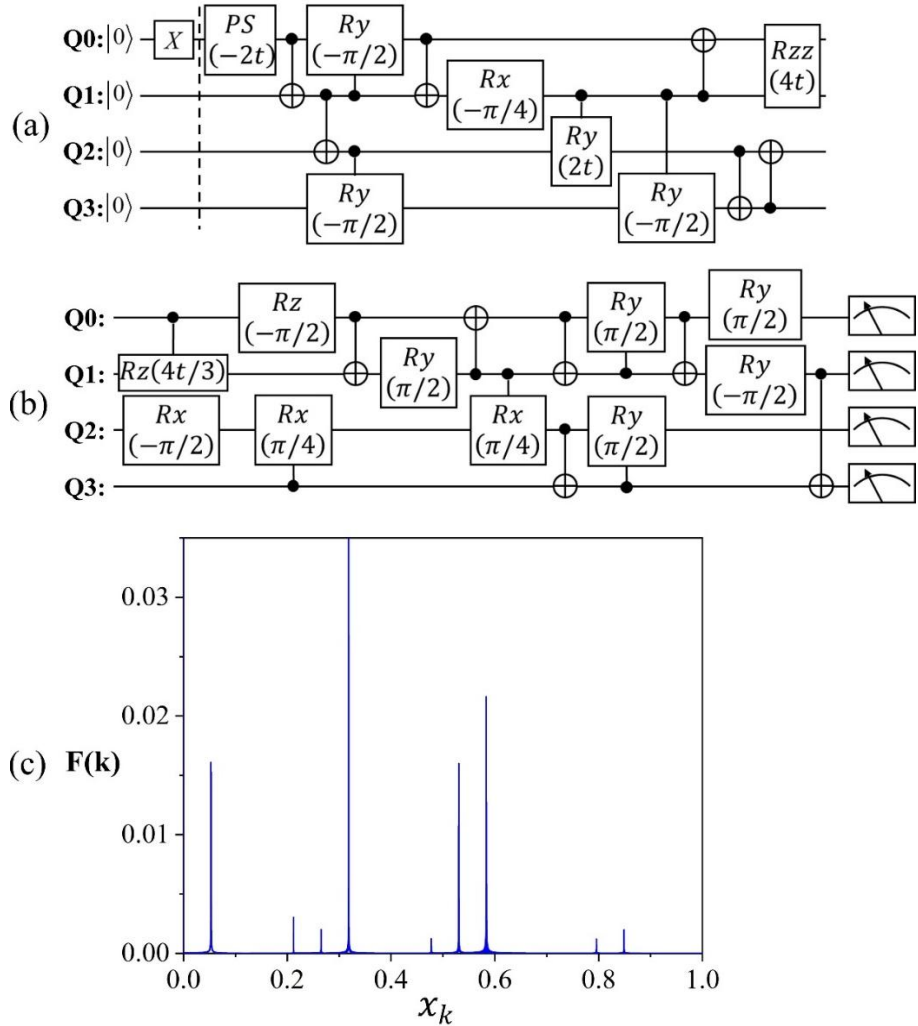

Figure 5. (a) and (b) depict the architecture of a four-qubit random quantum circuit used to test MQTE algorithm under circuit noise. (c) shows the spectrum of its corresponding $p_{ij}(n)$ spectrum under ideal conditions (i.e., without circuit noise or sampling errors).

To assess the resilience of the MQTE algorithm under realistic conditions, we simulate its performance in the presence of circuit noise modeled via a Pauli channel. Our analysis focuses on a four-qubit random quantum circuit, as shown in Figure

5(a)(b), which encodes the time evolution operator $e^{-i\hat{H}t}$ for some unknown Hamiltonian. The dashed line marks the preparation of the reference state $|\varphi_j\rangle = |0001\rangle$, achieved by applying an $X$ gate to qubit Q0. We set the total number of time steps $N = 10^4$ and sampling interval $\Delta = 0.5$. Figure 5(c) displays the ideal DFT spectrum of $p_{ij}(n)$ for the case where $|\varphi_i\rangle = |\varphi_j\rangle$, i.e., measuring the probability of returning to the initial state. Under perfect conditions (no circuit noise or sampling error), nine distinct spectral peaks are observed, corresponding to energy gaps between eigenstates. The most prominent peak occurs at $x_k = 0.3183$ with amplitude $F(k) = 0.04288$.

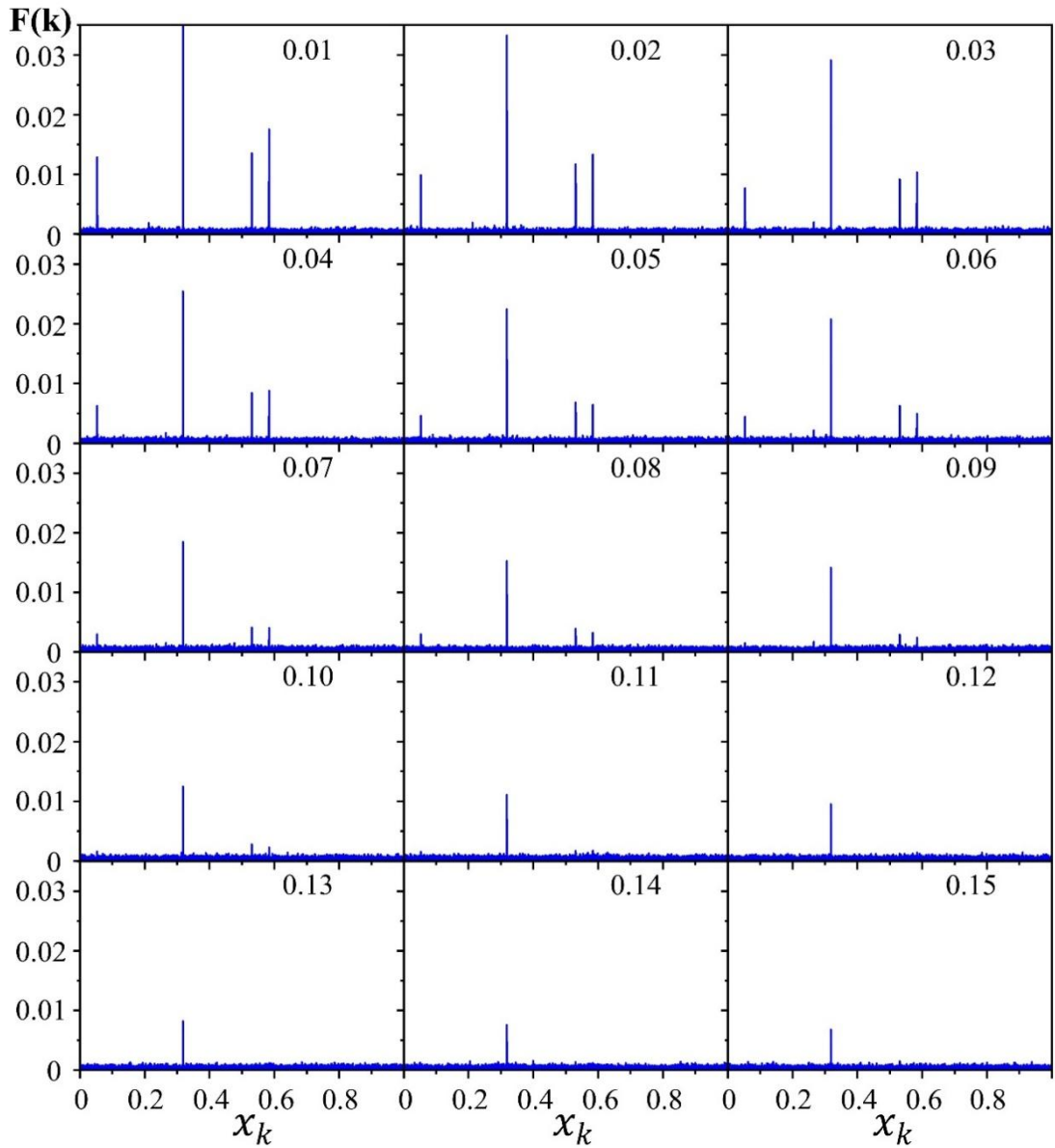

Figure 6 shows the DFT spectra of $p_{ij}(n)$ obtained by applying the MQTE algorithm to the 4-qubit random quantum circuit depicted in Figures 5(a)(b), under varying noise strengths $\gamma$ (ranging from 0.01 to 0.15). All simulation parameters are identical to those used in Figure 5(c), except that the number of measurements per time step is fixed at $M = 100$.

We now introduce circuit noise using a Pauli channel model, defined as:

$$\varepsilon(\rho) = (1 - \gamma)\rho + \frac{\gamma}{3}(X\rho X + Y\rho Y + Z\rho Z) \tag{20}$$

where $\rho$ is the density matrix of the quantum state, and $\gamma \in [0,1]$ denotes the noise strength. When $\gamma = 0$, the channel is identity; when $\gamma = 1$, each gate application is followed by a uniformly random Pauli error. This model captures the essence of gate infidelity due to control errors and decoherence.

Figure 6 shows the DFT spectra of $p_{ij}(n)$ under different levels of circuit noise ($\gamma = 0.01, 0.02, \ldots, 0.15$), with all other parameters identical to those in Figure 5(c), except that the number of measurements per time step is fixed at $M = 100$. As expected, both circuit noise and finite sampling contribute to a stochastic background in the spectrum. While this background masks weaker peaks, the positions of all resolvable spectral features remain unchanged compared to the noiseless case. This demonstrates that circuit noise does not distort the underlying energy spectrum—a key indicator of the algorithm's robustness.

However, unlike sampling noise, which primarily affects signal-to-noise ratio without altering signal amplitude, circuit noise systematically suppresses the height of all spectral peaks. As $\gamma$ increases, the amplitude of each peak decreases monotonically until it becomes indistinguishable from the noise floor. Specifically:

- At $\gamma \leq 0.08$, only 4 out of 9 peaks are detectable;
- At $0.09 \leq \gamma \leq 0.10$, only 3 peaks survive;
- At $\gamma \geq 0.11$, only the dominant peak at $x_k = 0.3183$ remains visible.

This behavior contrasts with sampling noise, which preserves relative peak heights but adds random fluctuations. Here, circuit noise introduces a global attenuation effect, reducing the overall signal strength while preserving spectral structure.

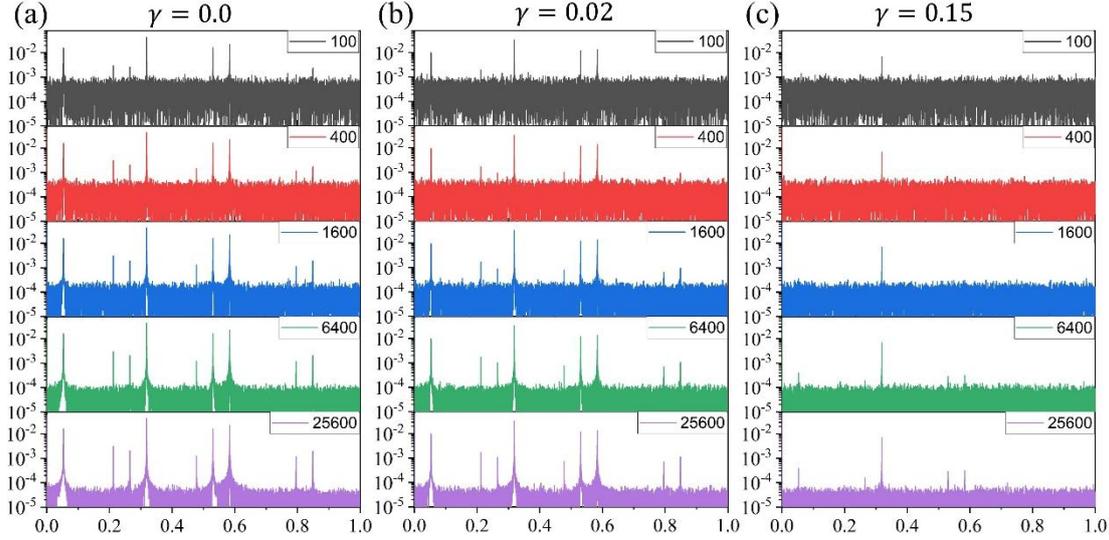

Figure 7 shows the DFT spectra of $p_{ij}(n)$ obtained by applying the MQTE algorithm to the 4-qubit random quantum circuit depicted in Figures 5(a)(b), under different noise strengths ($\gamma = 0.0, 0.02, 0.15$) and varying numbers of measurements per time step ($M = 100, 400, 1600, 6400, 25600$).

To investigate how sampling budget mitigates this degradation, we examine the impact of increasing $M$ under fixed noise levels. Figure 7 presents the DFT spectra for $\gamma = 0.02$ and $\gamma = 0.15$ across multiple values of $M$. A clear trend emerges: increasing $M$ reduces the noise floor proportionally to $M^{-1/2}$, allowing previously obscured peaks to reappear. For instance: At $\gamma = 0.15$, only one peak is visible at $M = 100$, but when $M = 25600$, five peaks become identifiable—more than in the weak-noise case $\gamma = 0.01$ with $M = 100$; At $\gamma = 0.02$, all nine peaks can be resolved even at $M = 1600$.

Importantly, the relative heights of the recovered peaks remain consistent across different noise regimes, indicating that the qualitative information about wavefunction overlaps—encoded in $c_{nj}c_{ni}c_{mj}c_{mi}$ —is preserved despite quantitative suppression. Thus, while circuit noise prevents accurate estimation of absolute overlap magnitudes, it does not destroy the relative hierarchy of spectral amplitudes. This allows for reliable identification of dominant excitations in the reference state, even under significant noise.

Our simulations results demonstrate that the MQTE algorithm possesses inherent robustness to circuit noise:

- Energy gaps are computed with quantitative precision—their positions are unaffected by noise.
- Wavefunction overlaps are estimated with qualitative accuracy—circuit noise

suppresses the amplitudes of all characteristic peaks with relative heights among distinct spectral peaks are preserved.
- The algorithm's resilience can be enhanced by increasing the sampling budget, trading off computational cost for noise tolerance.

**Theoretical Analysis**

To provide a theoretical foundation for these observations, we analyze how circuit noise affects the time-dependent signal $p_{ij}(n)$. Let $N_G$ denote the total number of quantum gates in the circuit implementing $e^{-i\hat{H}t}$. Each gate is subject to independent Pauli (or any other unitary or non-unitary random matrix) errors with probability $\gamma$. Define $f(\gamma) = (1-\gamma)^{N_G}$ as the probability that no error occurs during the entire evolution. Then, the measured collapse probability $\bar{\bar{p}}_{ij}(n)$ at time $n$ is given by:

$$\bar{\bar{p}}_{ij}(n) = f(\gamma)\bar{p}_{ij}(n) + [1 - f(\gamma)]\bar{q}_{ij}(n) \tag{21}$$

Where $\bar{p}_{ij}(n)$ is the noise-free signal, which follows $N(p_{ij}(n), \sigma_0^2/M)$, $\bar{q}_{ij}(n)$ represents the outcome under erroneous evolution.

We assume that under random Pauli noise, the output state collapses uniformly across all computational basis states, so $\mathbb{E}[q_{ij}(n)] = C_0$, a constant independent of $i, j, n$, with variance $Var[q_{ij}(n)] = \sigma_q^2$. After $M$ independent runs, the empirical estimator satisfies:

$$\mathbb{E}[\bar{\bar{p}}_{ij}(n)] = f(\gamma)p_{ij}(n) + [1 - f(\gamma)]C_0 \tag{22}$$

$$Var[\bar{\bar{p}}_{ij}(n)] = \frac{f^2(\gamma)\sigma_0^2 + [1-f(\gamma)]^2\sigma_q^2}{M} \tag{23}$$

where $\sigma_0^2$ is the variance of the clean signal and the covariance term vanishes due to independence. Taking the DFT, we find that the noisy spectrum consists of two components:

1. A scaled version of the true spectrum: $f(\gamma)DFT[p_{ij}(n)]$,

2. A DC offset: $[1-f(\gamma)]DFT[C_0]$, contributing only at zero frequency.

Thus, the frequency structure is preserved, but all non-zero-frequency peaks are attenuated by a factor $f(\gamma)$. This explains the monotonic decay of peak amplitudes with increasing $\gamma$ in Figure 6. Moreover, since $f(\gamma) \to 0$ exponentially with $N_G$, the signal becomes negligible for long circuits unless $\gamma \ll 1$.

Based on Eq.23, the total noise variance in the time domain scales as $M^{-1}$, but crucially, substituting $0 \leq f(\gamma) \leq 1$ into Eq.(23) yields that the upper and lower bounds of $Var[\bar{\bar{p}}_{ij}(n)]$ are $\sigma_0^2/M$ and $\sigma_q^2/M$, which do not increase infinitely as noise increases, so the relationship between variance and $\gamma$ is relatively insignificant. Therefore, as seen in Figure 7 above, the amplitude of the white noise remains approximately the same across different levels of circuit noise, and its intensity depends only on the number of samples.

The theoretical analysis is in agreement with the numerical simulations: circuit noise attenuates the amplitudes of genuine spectral peaks, rendering them more vulnerable to being obscured by the white noise floor induced by finite sampling. To recover weaker features in the spectrum, one must increase the number of measurements to suppress this sampling noise-effectively trading higher sampling cost for greater resilience against circuit-level errors. This trade-off is precisely what we observe in the simulations above. Based on this understanding, we can now qualitatively estimate the additional sampling overhead imposed by circuit noise. To ensure that intrinsic spectral peaks remain distinguishable over the white-noise floor, we require:

$$Var[\bar{\bar{p}}_{ij}(n)] \leq f(\gamma)p_{ij}(n) \tag{24}$$

Substituting Eqs. (21) and (26), we obtain:

$$\frac{\sigma^2}{M} \leq (1-\gamma)^{N_G} p_{ij}(n) \quad \Rightarrow \quad M = O\left[\left(\frac{1}{1-\gamma}\right)^{N_G}\right] \tag{25}$$

This reveals a sobering yet nuanced insight: although the MQTE algorithm can trade increased sampling overhead for resilience against circuit noise, the required number of measurements grows exponentially with the circuit depth $N_G$. For deep circuits—common in long-time evolutions or complex Hamiltonians—this sampling cost quickly becomes prohibitive. Fortunately, the severity of this exponential scaling is strongly moderated by the gate error rate $\gamma$. As quantum hardware continues to improve, $\gamma$ is expected to decrease steadily. Crucially, even modest reductions in $\gamma$ lead to dramatic gains in the maximum circuit depth that remains feasible within a fixed sampling budget. For instance, assuming a practical upper limit of $M = 10^4$ measurements per time step:

- At $\gamma = 10^{-2}$, the largest tolerable circuit depth is $N_G \approx 916$;
- At $\gamma = 10^{-3}$, it jumps to $N_G \approx 9,206$;
- At $\gamma = 10^{-4}$, it further increases to $N_G \approx 92,099$.

Thus, while current NISQ-era devices-with typical gate errors around $10^{-3}$ to $10^{-2}$ - may struggle to implement MQTE for highly complex or long-duration evolutions, ongoing advances in gate fidelity will substantially expand its applicability. We therefore anticipate that, as hardware matures and $\gamma$ continues to shrink, the MQTE approach will become increasingly viable for probing richer and more intricate quantum many-body systems.

## 4. Experimental Validation on Real Quantum Hardware

To further assess the practical viability of the MQTE algorithm, we implemented the MQTE algorithm on Tianyan-176-II, a superconducting quantum processor based on the Zuchongzhi-2 quantum chip, which supports up to 66 physical qubits[33–35]. The device exhibits typical gate error rates of 0.18% for single-qubit gates and 1.06% for two-qubit gates, with a readout error rate of 2.05%—representative of current NISQ-era hardware performance.

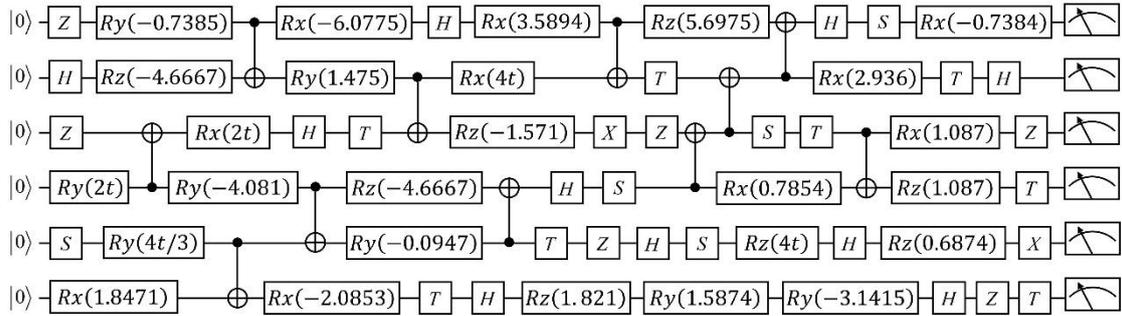

Figure 8. The architecture of a random quantum circuit used to test MQTE algorithm on practical quantum device.

To perform spectral estimation, we constructed a six-qubit random quantum circuit (as shown in Figure 8) that encodes the time evolution operator $e^{-i\hat{H}t}$ for a specific Hamiltonian. The reference state was prepared as $|\varphi_j\rangle = |000000\rangle$, and the time evolution was discretized using $N = 4000$ steps with a fixed time interval $\Delta = 0.5$. We present in Figure 9 the experimental DFT spectra of $p_{ij}(n)$ for four distinct measurement outcomes: $|000000\rangle$, $|000010\rangle$, $|010010\rangle$, and $|110110\rangle$. Each subplot compares the ideal noiseless spectrum (cyan) with experimental results obtained using 100 (red), 1000 (blue), and 10,000 (green) measurement shots per time step. Remarkably, despite substantial hardware imperfections—gate errors and limited coherence times—the positions of all dominant spectral peaks remain unchanged across all cases, demonstrating that MQTE preserves spectral accuracy even in the presence of realistic noise. Furthermore, increasing the shot count systematically suppresses the noise floor, revealing previously masked features and highlighting the effectiveness of sampling as a mitigation strategy.

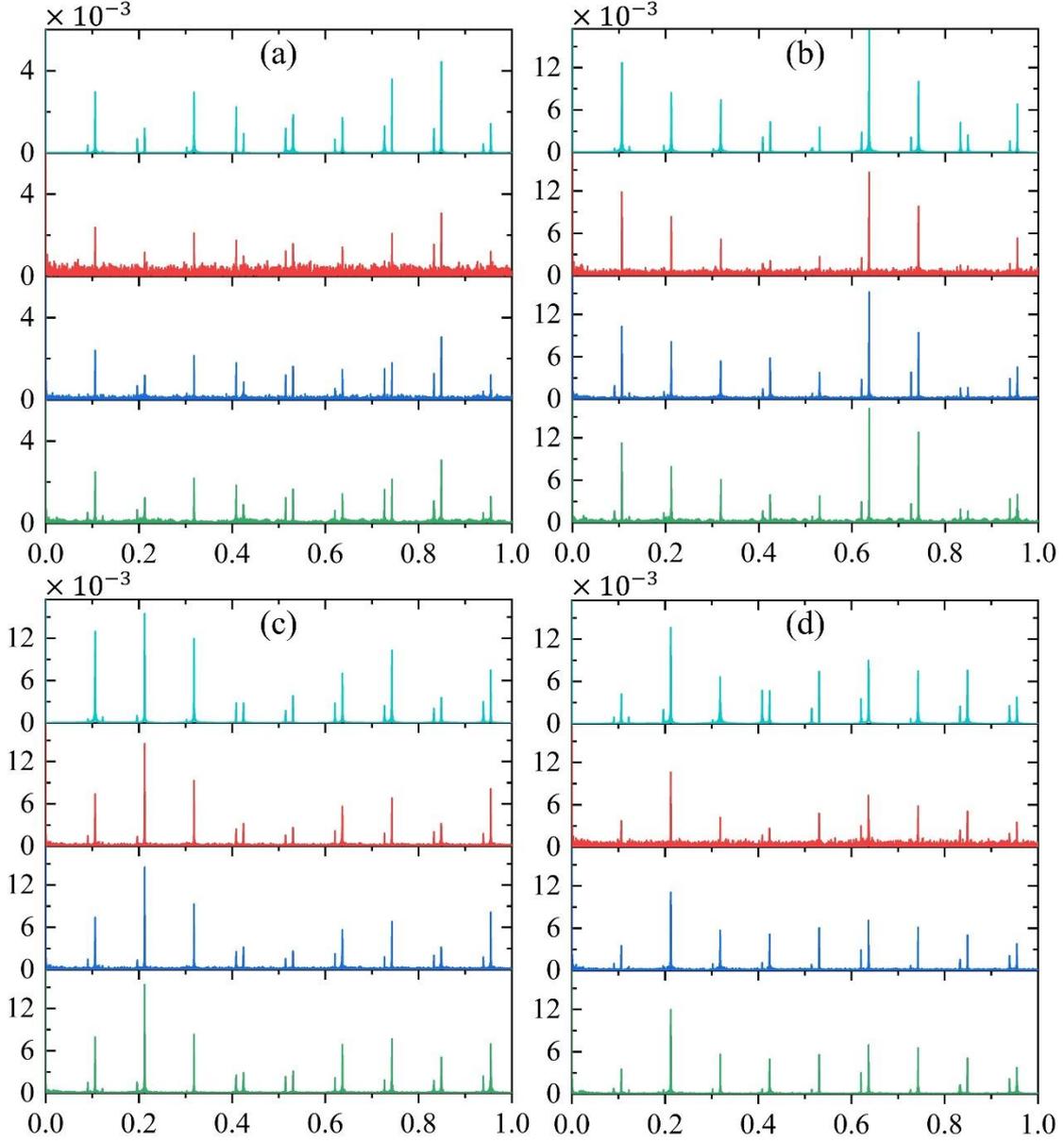

Figure 9. DFT spectra of $p_{ij}(n)$ for measurements collapsing into the states $|000000\rangle$(a), $|000010\rangle$(b), $|010010\rangle$(c), and $|110110\rangle$(d). In each panel, the topmost cyan curve shows the ideal simulation in the absence of sampling noise and circuit errors. The red, blue, and green curves correspond to experimental results obtained on the Tianyan-176-II quantum processor with 100, 1000, and 10,000 measurement shots per time step, respectively.

## 5. Conclusion

In this paper, we have proposed and systematically investigated the Measured

Quantum Time Evolution algorithm for extracting spectral information from quantum Hamiltonians. By evolving a reference state under real-time dynamics and performing projective measurements at discrete time steps, MQTE reconstructs energy gaps through classical spectral analysis of the recorded probability signals. The algorithm eliminates the need for ancillary qubits and controlled operations, significantly reducing circuit depth and gate count, making it well-suited for implementation on near-term quantum devices.

Theoretical analysis demonstrates that MQTE is inherently robust to both circuit noise and sampling errors, which predominantly manifests as a uniform attenuation of signal amplitudes and an additive white-noise background, without distorting the spectral peak positions that correspond to true energy gaps. Numerical simulations on 1D and 2D Heisenberg models confirm the algorithm's accuracy and clarify the roles of sampling interval and measurement budget in spectral resolution and signal-to-noise ratio. Experimental validation on the Tianyan-176-II superconducting quantum processor substantiates the practical viability of MQTE under realistic noisy conditions. Even in the presence of significant gate and readout errors, increasing the number of measurement shots enables clear recovery of spectral features, underscoring the algorithm's utility in noisy environments. Looking forward, MQTE's simplicity and noise resilience make it a promising candidate for studying excitation spectra, quench dynamics, and other spectral properties of complex quantum materials and chemical systems on existing quantum hardware. Future work will explore its integration with advanced error mitigation techniques and its application to larger, more physically relevant Hamiltonians.

## Acknowledgments

This work is supported in part by the National Natural Science Foundation of China (Grant Nos. 12504267, 22273069), the Natural Science Foundation of Hubei Province (Grant No. 2025AFB274), and the CPS-Yangtze Delta Region Industrial Innovation Center of Quantum and Information Technology-MindSpore Quantum Open Fund. The experimental implementation on real quantum hardware was enabled by the Tianyan Quantum Computing Cloud Platform, developed by China Telecom Quantum Information Technology Group Co., Ltd. The source code and experimental data supporting the findings of this study are openly available in the Gitee repository at

https://gitee.com/xie-qingxing/quantum-algorithm-mqte under the Apache License 2.0.